\documentclass[superscriptaddress,column,showpacs,a4paper,
amssymb,amsmath,nobibnotes,aps,prd,longbibliography,
showkeys,
nofootinbib,notitlepage]{revtex4-1}
\usepackage{bigints}
\usepackage{graphicx}
\usepackage{float}
\usepackage{epsf}
\usepackage{bm}
\usepackage{amsmath}
\usepackage{amsfonts}
\usepackage{amssymb}
\usepackage{epstopdf}
\usepackage{natbib}
\usepackage{color}%
\providecommand{\U}[1]{\protect\rule{.1in}{.1in}}
\newcommand{\be}{\begin{equation}}
\newcommand{\ee}{\end{equation}}

\newcommand{\mincir}{\raise
-3.truept\hbox{\rlap{\hbox{$\sim$}}\raise4.truept\hbox{$<$}\ }}
\newcommand{\magcir}{\raise
-3.truept\hbox{\rlap{\hbox{$\sim$}}\raise4.truept\hbox{$>$}\ }}

\newtheorem{remark}{Remark}[section]
\usepackage{xcolor}
\usepackage{hyperref}
\hypersetup{colorlinks=true, linkcolor=blue,citecolor=magenta}

\begin{document}

\title{{
Recovering Einstein’s Mature View of Gravitation: A Dynamical Reconstruction Grounded in the Equivalence Principle
}}

\author{Jaume de Haro}
\email{jaime.haro@upc.edu}
\affiliation{Departament de Matem\`atiques, Universitat Polit\`ecnica de Catalunya, Diagonal 647, 08028 Barcelona, Spain}

\author{Emilio Elizalde}
\email{elizalde@ice.csic.es}
\affiliation{Institut de Ciències de l'Espai, ICE/CSIC and IEEC, Campus UAB, C/Can Magrans, s/n, 08193 Bellaterra, Barcelona, Spain}

\begin{abstract}
The historical and conceptual foundations of General Relativity are revisited, putting the main focus on the physical meaning of the invariant $ds^2$, the Equivalence Principle, and the precise interpretation of spacetime geometry. It is argued that Albert Einstein initially sought a dynamical formulation in which $ds^2$ encoded the gravitational effects, without invoking curvature as a physical entity. The now more familiar geometrical interpretation—identifying gravitation with spacetime curvature—gradually emerged through his collaboration with Marcel Grossmann and the adoption of the Ricci tensor in 1915. Anyhow, in his 1920 Leiden lecture, Einstein explicitly reinterpreted spacetime geometry as the state of a physical medium—an “ether” endowed with metrical properties but devoid of mechanical substance—thereby actually rejecting geometry as an independent ontological reality.

Building upon this mature view, gravitation {is}  reconstructed  from the {Weak} Equivalence Principle, understood as the exact compensation between inertial and gravitational forces acting on a body {under a uniform gravitational field}. From this fundamental principle, {together with an
extension of Fermat's
Principle to massive objects}, the invariant $ds^2$ is obtained, first in the static case, where the gravitational potential modifies the flow of proper time. Then, by applying the Lorentz transformation to this static invariant, its general form is derived for the case of matter in motion. The resulting  invariant reproduces the relativistic form of Newton’s second law in proper time and coincides with the weak-field limit of General Relativity in the harmonic gauge.

This approach restores the operational meaning of Einstein’s theory: spacetime geometry represents dynamical relations between physical measurements, rather than the substance of spacetime itself. By deriving the gravitational modification of the invariant directly from the {Weak} Equivalence Principle, {Fermat Principle} and Lorentz invariance, this formulation clarifies the physical origin of the metric structure and resolves long-standing conceptual issues—such as the recurrent hole argument—while recovering all the empirical successes of General Relativity within a coherent and sound Machian framework.

\end{abstract}

\vspace{0.5cm}


\maketitle

\section{Introduction}

Since its publication in 1915, Einstein’s theory of General Relativity (GR) has been celebrated as one of the greatest intellectual achievements in the history of physics. Its mathematical elegance and predictive power have inspired generations of physicists and mathematicians who, on top of formidable applications of the theory to describe the world, do not cease to praise its conceptual beauty. Yet, alongside these many successes, the theory has also given rise to enduring conceptual ambiguities. One of the most persistent misconceptions is that spacetime itself constitutes a physical substance—something that can literally be “bent” or “curved” by matter and energy. Such imagery, often reinforced by popular accounts and some professional expositions, too, can obscure the operational role of the mathematical structures that underpin the theory.

Einstein himself, in spite of being quite cautious, contributed to some of these ambiguities. In 1907, when he formulated the Equivalence Principle \cite{Einstein1907}, he introduced the heuristic idea that a homogeneous gravitational field could be  replaced by a uniformly accelerated frame. In fact, this idea was extremely useful to him at the very beginning, when he was starting to construct the theory, as he later recognized on several occasions. In 1911, he applied this principle to light, thereby predicting the gravitational redshift and the deflection of light rays \cite{Einstein1911}. By 1915, he had finally completed the field equations, in which gravitation was represented through the metric tensor and the curvature of spacetime \cite{Einstein1915}. Finally, in his 1920 Leiden lecture {\it Ether and the Theory of Relativity} \cite{Einstein1920}, he reinterpreted the metric field as a new kind of “ether”—not the mechanical medium of the nineteenth century, but the metrical structure that determines the behavior of rods and clocks under the influence of gravity. This “ether,” Einstein emphasized, could not be decomposed into parts or set into motion, yet it embodied the physical state of spacetime.

It happened, however, that subsequent expositions often extended these heuristic notions into literal ontological claims, fostering a persistent reification of geometry. Over time, the image of curved spacetime came to be treated almost as a material entity. The rise of differential geometry and tensor calculus reinforced the belief that gravitation was geometry itself, resulting in a complete mathematicalization of physics, at the very least, of gravity as a spacetime theory. Classic presentations of the theory by Weyl \cite{Weyl1918}, Eddington \cite{Eddington1923}, and Pauli \cite{Pauli} popularized the idea that matter dictates the curvature of spacetime, which in turn governs the motion of matter—an aphorism later echoed by John Archibald Wheeler, with enormous success \cite{TW}. While mathematically elegant, as an interpretation, at first sight, of Einstein's field equations, such imagery risks obscuring the deep physical content of the theory.

This conceptual tension was already evident in Einstein’s time. Nikola Tesla, for instance, rejected General Relativity on the grounds that “space cannot be curved, for the simple reason that it can have no properties” \cite{Tesla}. In the Soviet Union, by contrast, skepticism toward GR often arose from dialectical materialism, which demanded a tangible substrate for all physical phenomena. Yet, some Soviet physicists—notably Lev Landau \cite{Landau} and Vladimir Fock \cite{Fock}—recognized the empirical success of GR and sought to reinterpret its concepts within a more physically grounded field-theoretical framework.

{This historical experience highlights a recurrent danger: the easy identification of pedagogical metaphors with physical reality. A commonly used graphical analogy invites us to imagine a large mass placed on an elastic mattress, which deforms under its weight, so that a small ball rolling nearby follows a curved trajectory. The image is meant to suggest, in a visually intuitive way, how motion can deviate from straight lines without invoking a force acting at a distance.

However, it must be emphasized that this picture is purely pedagogical. The deformation of the mattress is caused by the weight of the mass—namely, by the very Newtonian gravitational force exerted by the Earth—and therefore does not literally represent what the theory asserts. In short, attempting to refute Newtonian gravity while implicitly relying on its own assumptions is conceptually circular, much like trying to prove the existence of God through Aquinas’s five ways while already presupposing that existence. Such reasoning may appear elegant, but in practice it obscures rather than clarifies the physical content of the theory.

For this reason, curvature should not be interpreted as the bending of a physical medium, but as a powerful mathematical tool for encoding gravitational effects. It must be understood operationally, in terms of measurable relations between clocks, rods, and light signals, rather than as a literal geometric deformation of space itself.

}

The aim of this paper is to revisit these interpretive issues and to recover the physical meaning that Einstein ultimately assigned to the theory. We argue that General Relativity should not be read primarily as a theory about the geometry of a material substratum, but as a dynamical framework governing how physical measurements transform under the influence of gravitation. By re-examining Einstein’s writings—starting from the 1907 formulation of the Equivalence Principle and ending in the 1920 Leiden lecture—we clarify how the modern geometric reading of GR emerged and how it can be reinterpreted in purely dynamical terms.

In particular, we show that a physically grounded perspective—at least in the weak-field regime relevant for solar-system dynamics—can be formulated without invoking curvature as an ontological entity. Rooted in the operational meaning of the Equivalence Principle, {the extension of Fermat’s principle \cite{Fermat} to massive bodies—according to which free trajectories extremize proper time—} and the Lorentz invariance of the laws of motion, this approach reconstructs gravitation as a dynamical manifestation of the interaction between matter and inertia.

In summary, our aim is in no way to try to deny the utility of the geometric formalism, but rather to restore its operational meaning. The geometrical structure of spacetime should be understood as a representation of dynamical relations among measurable quantities—times, lengths, and accelerations—rather than as a physical substance endowed with curvature. At least in the weak-field limit, General Relativity can thus be viewed as a dynamical law governing the mutual relations between matter, inertia, and the propagation of signals (much closer, in a way, to the special theory of relativity), rather than as a statement about the material properties of spacetime itself. This viewpoint is consistent with Einstein’s mature interpretation presented in Leiden, where he emphasized that the “ether” of General Relativity is not a mechanical medium, but a system of metrical relations that encode how rods and clocks behave under gravity.

\section{From the Equivalence Principle to General Relativity: The Historical Genesis of Einstein’s Theory}

To understand the conceptual tensions underlying General Relativity, it is essential to revisit the sequence of insights and assumptions that guided Einstein from the Equivalence Principle to the generally covariant field equations. This path did not begin as a geometrical construction, but as a physical search for laws compatible with the Principle of Relativity and the empirical equivalence between inertial and gravitational mass.

We will start with Minkowski’s seminal (an extremely important) reformulation of Einstein’s Special Theory of Relativity (STR) \cite{Minkowski,Einstein1905}, where he introduced the invariant quantity
\begin{equation}
ds^2 = dt^2 - d{\bf x}\cdot d{\bf x},
\end{equation}
which remains unchanged under Lorentz transformations. This expression unified space and time into a single four-dimensional continuum, endowing the theory with a new mathematical elegance. However, its physical meaning is more concrete than mere geometric abstraction. As Minkowski already emphasized, the integral
\begin{equation}
s = \int ds,
\end{equation}
taken along a world-line between two events, represents the proper time experienced by an observer moving along that path. The differential element $ds$ therefore corresponds to the infinitesimal proper time measured by a clock carried by the observer.

This interpretation placed temporal experience—not spatial geometry—at the heart of relativity: motion and dynamics could now be expressed as relations between locally-measured intervals of time. When Einstein later sought to extend the Principle of Relativity to accelerated systems, he naturally reinterpreted $ds$ as a dynamical quantity that should encode the effects of gravitation, too. From this standpoint, the transition from Special to General Relativity can be seen as the progressive generalization of the invariant $ds$, from a flat Minkowskian form to one that already accounts for the influence of matter and acceleration.

In fact, this invariant had already been introduced by Planck \cite{Planck} in a different context and without an explicit physical interpretation. Planck demonstrated that, in the absence of gravitation, the equations of motion of free particles in Special Relativity follow from the variational principle
\begin{equation}
\delta \int ds = 0,
\end{equation}
that is, by extremizing the proper time along the trajectory of the particle.

We now turn to Einstein’s attempt to incorporate gravitation into the theory of Special Relativity. In 1907, he formulated the first version of the Equivalence Principle, according to which there exists,  a complete physical equivalence between a uniformly accelerated reference frame and a system at rest in a homogeneous gravitational field.

This assumption entails two main consequences. First, by identifying uniformly accelerated motion with rest state in a homogeneous gravitational field, Einstein argued that the principle of relativity could be extended beyond inertial frames to include uniformly accelerated ones. Implicit in this step is the conviction that Special Relativity remains valid, at least approximately, for observers at rest in a uniform gravitational field—a situation analogous to that experienced on the Earth’s surface. Second, the Equivalence Principle allows for a uniform gravitational field to be replaced by a uniformly accelerated frame, which is far more tractable theoretically.

As a practical application, Einstein showed that the flow of time in clocks is affected by gravity: a clock placed at a weaker gravitational potential runs faster. A similar heuristic  conclusion can be reached more directly by combining the local Lorentz transformation with Newton’s second law. 

For an infinitesimal Lorentz boost, disregarding the Lorentz factor since it is of the second order in the velocity, one has approximately  
$$\Delta t' \cong \Delta t - v\,\Delta x.$$  
For uniformly accelerated motion, \(v \cong a\,\Delta t\), where \(a\) denotes acceleration. Substituting, one gets  
$$\Delta t' \cong \Delta t(1 - a\,\Delta x).$$  
And using the Newtonian relation \(a = -\partial_x \Phi\), or \(\Phi = -a x\) for constant acceleration, one obtains  
$$\Delta t' \cong \Delta t(1 + \Delta\Phi),$$  
which expresses that the rate of time depends on the gravitational potential. This result is formally equivalent to the expression derived by Einstein in 1907, where he found that a clock located at potential \(\Phi\) runs faster by a factor \(1+\Phi\) relative to another placed at the origin.

{This argument shows that the gravitational redshift follows directly from Lorentz invariance combined with the Equivalence Principle, without invoking any assumption about spacetime curvature. For completeness, the same result can also be obtained in a purely non-relativistic manner from the classical Doppler effect, as follows.

Let us consider the classical Doppler formula:
\begin{eqnarray}
\nu = \left(\frac{v + v_r}{v + v_f}\right)\nu_0,
\end{eqnarray}
where $\nu$ is the observed frequency, $\nu_0$ the emitted frequency, $v$ the wave velocity, $v_r$ the velocity of the receiver, and $v_f$ the velocity of the source.

For light ($v \equiv 1$ in units where $c=1$) and with the source at rest, this expression reduces to
\begin{eqnarray}
\nu = \left(1 + v_r\right) \nu_0.
\end{eqnarray}

If the receiver undergoes a constant acceleration $a$ along the $OZ$ direction, its velocity at time $t$ is
$v_r = a t = a h$, where $h$ is the height at which the receiver is located. One then obtains
\begin{eqnarray}
\nu = \left(1 + a h\right) \nu_0.
\end{eqnarray}

Applying the Equivalence Principle allows one to replace $a h$ by a homogeneous gravitational potential $\Phi$, yielding the approximate expression for the gravitational frequency shift:
\begin{eqnarray}
\nu = \left(1 + \Phi\right) \nu_0.
\end{eqnarray}

This derivation is strictly non-relativistic and valid only for small velocities and weak fields. Nevertheless, it provides a particularly transparent illustration of the connection between acceleration, gravitation, and frequency shift, bypassing the full relativistic formalism employed in Einstein’s 1911 analysis.

}

\

{By 1911, invoking the Equivalence Principle once again, Einstein argued that a ray of light passing near a massive body should follow a curved trajectory, since the propagation of light is subject to the same gravitational influence as material bodies. Although his result amounted to only half of the value later predicted by General Relativity, it nevertheless constituted the first theoretical prediction of the gravitational deflection of light.

}

In 1912, Einstein extended the principle of relativity to uniformly accelerated frames. He assumed that local measuring rods retain their length under uniform acceleration, so that Euclidean rules for spatial measurement remain valid locally. Simultaneously, he recognized that the passage of time for clocks is affected by acceleration—or equivalently, by a homogeneous gravitational field—so that temporal intervals must be treated differently. This framework allowed him to explore the effects of gravity on physical processes while preserving a well-defined notion of local spatial geometry, thus bridging the gap between accelerated frames and the emerging concept of gravitational fields.

{Specifically, in \cite{Einstein1912b,Einstein1912a}
Einstein considers a uniformly accelerated system, which we shall call
$\Sigma_1,$
with coordinates
$(t,{\bf x}),$
which accelerates, with acceleration $a$, along the direction of the $OX$ axis with respect to a system
$\Sigma_2$
with coordinates
$(T,{\bf X})$,
which represents the spacetime of special relativity (without gravity). In this inertial system, the spacetime interval retains the Minkowskian form:
$$
ds^2 = dT^2 - d{\bf X}\cdot d{\bf X}.
$$

On the other hand, Einstein assumes that, for an observer in the accelerated system
$\Sigma_1$,
the invariant takes the form
\begin{eqnarray}
ds^2 = c^2(x)\, dt^2 - d{\bf x}\cdot d{\bf x},
\end{eqnarray}
that is, the acceleration of the system is reflected in a modification of the coefficient multiplying $dt^2$, which he interprets as a spatial variation of the speed of light. By contrast, the spatial part of the interval remains Euclidean. Today we know that this assumption is incomplete: in a genuine gravitational field, modifications of the spatial term also appear. However, it should be emphasized that in 1912 Einstein was advancing through uncharted territory; these papers are preliminary attempts, still far from the mature geometric formulation that he would develop between 1913 and 1915.

Einstein uses his physical intuition and assumes that, for small times $t$, the relation between coordinates is of the form
\begin{eqnarray}
Y = y, \qquad Z = z, \qquad
X = \lambda + \alpha t^2 + \cdots, \qquad
T = \beta + \gamma t + \delta t^2 + \cdots.
\end{eqnarray}

Substituting into the invariant and comparing coefficients, one finds that the transformation, up to order $t^2$, must be
\begin{eqnarray}
X = x + \frac{a\, c(x)}{2}\, t^2, \qquad
Y = y, \qquad Z = z, \qquad
T = c(x)\, t,
\end{eqnarray}
with
$
c(x) = 1 + a x$.

Now, using the Equivalence Principle and identifying $a x$ with a uniform gravitational field $\Phi$, the invariant takes the form
\begin{eqnarray}
ds^2 = (1 + \Phi)^2 dt^2 - d{\bf x}\cdot d{\bf x}\cong (1 + 2\Phi)dt^2 - d{\bf x}\cdot d{\bf x}.
\end{eqnarray}

Therefore, on the basis of this result, Einstein assumed that in the case of a static gravitational field the invariant $ds$—which he had not yet explicitly associated with proper time—must take the form}
\begin{equation}\label{invariant}
ds^2 = c^2({\bf x}) dt^2 - d{\bf x}\cdot d{\bf x},
\end{equation}
where $c({\bf x})$ is an unknown function related to the gravitational potential. More importantly, after analyzing the case of a uniform gravitational field, Einstein realized that the dynamics of particles in a general static potential could still be derived from the same variational principle underlying Special Relativity, namely \cite{Einstein1912b}:
\begin{equation}
\delta \int ds = 0.
\end{equation}

This meant that the motion of a freely falling body could be described as an extremal trajectory of the invariant $ds$, just as in Minkowski space-time, but now in the presence of gravity. In other words, introducing a gravitational potential did not require abandoning the variational framework of STR; it merely modified the expression of $ds$. This insight suggested that gravity could be encoded in the structure of the invariant itself—constituting the seed of a dynamical interpretation of gravity, independent of geometric curvature—what paved the way to the later formulation of General Relativity.

{The next step was to find the relationship between $c({\bf x})$ and the gravitational potential. Motivated by the classical Poisson equation, Einstein proposed
\begin{equation}\label{E1}
\Delta c = \kappa c \rho,
\end{equation}
where $\kappa$ is a universal constant and $\rho$ the matter density.

However, it soon became clear that the dynamical equations of motion together with (\ref{E1}) led to a serious inconsistency: they violated momentum conservation.

To see this, recall that for the Poisson equation $\Delta \Phi = \kappa \rho$, one can define a stress-like tensor
\begin{equation}
\mathfrak{W} = d\Phi \otimes d\Phi - \frac12 |\nabla \Phi|^2 (dx+dy+dz) \otimes (dx+dy+dz),
\end{equation}
so that the force density ${\bf F} = -\rho \nabla \Phi$ satisfies
\begin{equation}
{\bf F} = - \frac{1}{\kappa} \mathrm{div}_{(3)} \mathfrak{W}, \quad \text{and thus,  by Gauss' theorem} \quad \int_{\mathbb{R}^3} {\bf F} \, dV = 0.
\end{equation}

In contrast, for (\ref{E1}) the force density is ${\bf F} = -\rho \nabla c$, and one finds
\begin{equation}
\int_{\mathbb{R}^3} {\bf F} \, dV = -\frac{1}{\kappa} \int_{\mathbb{R}^3} \frac{\Delta c}{c} \nabla c \, dV \neq 0,
\end{equation}
so no global momentum conservation law of the same form exists.

Since conservation of energy and momentum was fundamental for Einstein, he attempted to modify his field equation in a nonlinear way to ensure that the total momentum of matter plus the gravitational field would be conserved. Nevertheless, these modifications introduced new conceptual and mathematical problems, and the resulting formulation remained inconsistent, particularly with the Equivalence Principle. After several attempts, Einstein abandoned this approach, recognizing that a more radical reformulation was necessary \cite{Girard}.

}

\subsection{The collaboration with Marcel Grossmann}

In 1913, Einstein began collaborating with mathematician Marcel Grossmann in the search for the field equations of General Relativity. Einstein’s guiding idea was to extend the validity of Special Relativity to all reference systems. To this end, he introduced the Principle of Covariance \cite{Norton}, which required that the field equations retain the same form in all coordinate systems—or, in Einstein’s language, that they transform covariantly. This demand could be fulfilled by treating the problem geometrically: regarding space-time as a pseudo-Riemannian manifold and constructing the equations from tensors, primarily bilinear forms.

From a modern perspective, it is also possible to build equations that are intrinsic to the manifold (independent of the coordinates) without necessarily restricting oneself to tensors. Such equations may not transform as tensors under changes of coordinates but can still represent genuine geometric invariants. This distinction highlights a subtle difference between Einstein’s original understanding of covariance and the broader intrinsic formulation available in differential geometry.

Following this geometric viewpoint, Einstein considered $ds$ as the “naturally measured” distance between nearby space-time points. This contrasts with Minkowski’s perspective, where $ds$ was primarily understood as the infinitesimal proper time.

Einstein and Grossmann outlined the integration of gravity into STR in the ``Entwurf paper'' \cite{Grossmann}, in an attempt to find a field equation generalizing the classical Poisson equation, namely
\begin{equation}
\frak{U} = \kappa \frak{T},
\end{equation}
where $\kappa$ is a universal constant related to gravitation, $\frak{T}$ is the stress tensor, and $\frak{U}$ is a tensor depending on the metric $\frak{g}$, whose entries are unknown functions.

As a candidate, they considered the Ricci tensor, but rejected it because, according to Grossmann, in the static weak-field limit it did not reduce to the Laplacian of the Newtonian potential. Modern analysis shows this claim to be incorrect, though the authors provided no explicit calculation. It has been conjectured that their use of (\ref{invariant}), interpreted geometrically as implying spatial flatness, may have been the main reason \cite{Straumann,Renn2007,Norton1984}.

{We can see this explicitly as follows. In Section~\ref{Ricci} we derive the exact form of the Ricci tensor in harmonic gauge, showing that it is precisely in this gauge that the correct Newtonian limit is obtained. However, the invariant (\ref{invariant}) does not satisfy the harmonic gauge condition, and therefore the Laplace equation cannot be recovered.}

Since no satisfactory candidate for $\frak{U}$ was found—at least one containing second-order derivatives in the spirit of the Newton–Poisson equation—Einstein, though open at the time to the possibility that $\frak{U}$ might involve higher-order derivatives, considered any such attempt premature. Consequently, he abandoned the requirement of general covariance and restricted the field equations to a narrower class of coordinate transformations.

The motivation for this restriction was the well-known hole argument \cite{Stachel2014,Iftime-Stachel,Earman-Norton,Weatherall}, according to which a fully covariant theory—particularly one involving the Ricci tensor—appears to lack a unique determination of the metric. In modern terms, Einstein’s equations, expressed through the Ricci tensor, are invariant under active hole diffeomorphisms: smooth one-to-one mappings that relocate the points of the manifold within an empty region (the “hole”) while dragging along all tensor fields \cite{Davis1,Haro2024}.

This feature conflicted with Einstein’s Machian conviction \cite{Mach} that the distribution of matter should uniquely determine inertia—and thus fix the components of the fundamental form $\mathfrak{g}$ on the pseudo-Riemannian manifold representing spacetime, which govern the motion of test particles.

Although the Entwurf equations were not correct, the paper made several important contributions. For instance, for a matter incompressible fluid, Einstein derived the conservation of the stress tensor from the dynamics of test particles obtained via the variational principle ($\delta \int ds$). {In modern terms \cite{Marsden,Comer}, the conservation of the stress–energy tensor can be expressed as
\begin{equation}\label{stress-tensor}
\mathrm{div}(\mathfrak{T}) = 0 
\quad \Longleftrightarrow \quad 
\mathrm{div}(\rho {\bf u}) = 0 
\quad\text{and}\quad 
\frac{D{\bf u}}{ds} = 0,
\end{equation}
where ${\bf u}$ denotes the four-velocity field. The condition $\mathrm{div}(\rho {\bf u}) = 0$ represents mass conservation, while the equation $D{\bf u}/ds = 0$ explicitly states that fluid elements (or test particles) move along geodesics, that is, in the absence of external non-gravitational forces. Both the divergence operator $\mathrm{div}$ and the covariant derivative $D/ds$ are defined with respect to the Levi-Civita connection compatible with the fundamental form $\mathfrak{g}$.

For completeness, recall that the four-divergence of a vector field ${\bf u}$ is defined as the trace of the endomorphism of the tangent space that maps any four-vector ${\bf w}$ to the covariant derivative $\nabla_{\bf w}{\bf u}$, namely
\begin{eqnarray}
    \mathrm{div}({\bf u}) \equiv \mathrm{Tr}\big({\bf w} \mapsto \nabla_{\bf w}{\bf u}\big).
\end{eqnarray}
Likewise, the four-gradient of a scalar function $p$ is introduced via the sharp operator as
\[
\mathrm{grad}(p) \equiv dp^{\sharp}, 
\qquad
\text{with} \qquad 
\mathfrak{g}\big(dp^{\sharp},{\bf w}\big) = dp({\bf w})
\quad 
\text{for all four-vectors } {\bf w}.
\]

}

Another key aspect was the derivation of field equations compatible with stress tensor conservation, analogous to the conservation law in electrodynamics.

\subsection{The final equations}

Einstein was never fully satisfied with the \textit{Entwurf} equations, particularly after discovering that they failed to account for the observed precession of Mercury’s perihelion. This discrepancy, together with the lack of full covariance, convinced him that the underlying structure of the theory was incomplete.
He therefore returned to the Ricci tensor—an object that, as we have already noted, he had previously considered and dismissed—as a potential foundation for the gravitational field equations.

Before adopting it, however, Einstein had to resolve his \textit{hole argument}, which seemed to imply a breakdown of determinism under general covariance: if the field equations were generally covariant, then two distinct metric fields related by a diffeomorphism would both satisfy the same matter distribution, apparently violating uniqueness. Initially, Einstein viewed this as a fatal flaw in general covariance. By late 1915, however, he realized that diffeomorphic solutions must represent the same physical situation—merely different  descriptions of one and the same reality. This insight not only restored general covariance but also preserved, in a new and more subtle form, Mach’s idea that the distribution of matter determines the inertial structure of space.

From the modern perspective, this means that General Relativity does not determine a single {metric field} (the fundamental form $\frak{g}$), but an entire equivalence class of solutions related by diffeomorphisms. The physical content resides not in any particular representative of this class, but in the invariants common to all of them. In this sense, General Relativity is truly a gauge theory: the diffeomorphism group plays a role analogous to the gauge group in electrodynamics, where potentials related by a gauge transformation describe the same electromagnetic field. To perform explicit calculations, however, one must select a particular representative (effectively a gauge choice) within this class. 

\begin{remark}
In the weak-field regime, linearization of the field equations around Minkowski space makes this analogy precise: infinitesimal diffeomorphisms correspond to gauge transformations of the perturbation, and the harmonic (or de Donder) gauge provides the most natural choice \cite{Fock}. It is precisely in this gauge that the equations reduce to their Newtonian limit, ensuring consistency with classical mechanics and clarifying the dynamical origin of the gravitational potential within the relativistic framework.
\end{remark}

Once covariance was secured, Einstein faced up the task of identifying the correct dynamical law. His first proposal was
\begin{equation}
\frak{Ric} = \kappa \frak{T},
\end{equation}
where $\frak{Ric}$ is the Ricci tensor and $\frak{T}$ the stress–energy tensor.
This form worked for electromagnetic fields, whose energy–momentum tensor is traceless, but failed for ordinary matter: the nonzero trace (Laue’s invariant) implied a violation of energy–momentum conservation, contradicting one of Einstein’s core principles.

In search of consistency, Einstein temporarily adopted the so-called \textit{unimodular condition} $\det(\frak{g})=-1$, which simplified the structure of the Ricci tensor and allowed him to test different covariant combinations. 
 {Regarding its physical interpretation, the unimodular condition
does not correspond, in this context, to a measurable constraint on physical
systems. There is no known physical situation in which fixing the determinant of
the metric would have an invariant operational meaning. Its role here is purely
instrumental.

}

Through successive refinements, he arrived at the expression presented on 25 November 1915 \cite{Einstein1915}:
\begin{equation}\label{E2}
\frak{Ric} = \kappa \left(\frak{T} - \frac{1}{2}\frak{g}T\right),
\end{equation}
where $T$ is the trace of $\frak{T}$.

This formulation differs only in rearrangement from the modern version,
\begin{equation}\label{E3}
\frak{Ric} - \frac{1}{2}\frak{g}R = \kappa \frak{T},
\end{equation}
where $R$ is the Ricci scalar. In this form, the contracted Bianchi identities \cite{Bianchi} do guarantee that the covariant divergence of the stress–energy tensor vanishes,
\begin{equation}
\mbox{div}(\frak{T}) = 0,
\end{equation}
ensuring local energy–momentum conservation.

It is worth noting that Einstein’s reasoning proceeded in the opposite direction: conservation of energy and momentum was not a consequence but a guiding principle. As he emphasized in his 1916 review \cite{Einstein1916}, the total energy–momentum—including that of the gravitational field—must be conserved. This requirement led him to modify his initial equation $\frak{Ric} = \kappa\frak{T}$ towards the balanced form (\ref{E2}), where the trace term precisely compensates the divergence of $\frak{Ric}$. In this sense, the final field equations can be read as expressing a dynamic equilibrium between matter and inertia rather than a purely geometrical postulate. Geometry thus emerges as the mathematical representation of this balance—a formal language encoding the mutual determination of gravitation and inertia within a generally covariant framework.

\section{Geometric Interpretation of General Relativity}

One of the most influential and enduring features of Einstein’s theory is its interpretation as a geometrical theory of gravitation. In this framework, gravity ceases to be a force in the Newtonian sense—a direct interaction acting at a distance—to become instead a manifestation of the curvature of a four-dimensional pseudo-Riemannian manifold. The motion of free particles is then determined not by an external force, but by the geometry itself: they follow geodesics defined by the metric tensor, i.e., by the first fundamental form of the manifold,  $\mathfrak{g}$, which encodes the distribution of matter and energy through Einstein’s field equations. This interpretation represents one of the most profound conceptual transformations in the history of physics, turning a dynamical interaction among bodies into a structural property of spacetime.

Anyway, this elegant geometrical picture was not Einstein’s starting point. Historically, his path toward General Relativity was guided not by geometry but by physical principles: the Equivalence Principle, the conservation of energy and momentum, and the requirement that the laws of physics be covariant under general coordinate transformations. Geometry entered only as a language—an indispensable mathematical framework—through which these physical ideas could be consistently expressed. It was during his collaboration with Marcel Grossmann in 1912–13 that Einstein first encountered the work of Riemann, 
Christoffel,  Ricci, and Levi-Civita, what lead to the Entwurf theory. As he later recalled in 1922 \cite{EinsteinAutobio}, 
\begin{quote}
{\it “Grossmann introduced me to the work of Riemann, Ricci and Levi-Civita. Without him, I would not have found my way”}.
\end{quote}

When the Entwurf equations revealed their limitations—in particular, their failure to achieve full covariance and to reproduce the Newtonian limit satisfactorily—Einstein returned to the geometric objects he had initially set aside. In a letter to Sommerfeld (28 November 1915), after deriving the final form of the field equations, he wrote:
\begin{quote}
{\it “The key was the realization that only the Ricci tensor could fulfill the physical requirements”}.
\end{quote}

Thus, the adoption of the Ricci tensor in late 1915 marked the convergence of Einstein’s physical reasoning with the geometric formalism that could embody it. The geometrical interpretation of gravitation was therefore not the origin but the culmination of his search for a dynamical and covariant description of inertia and gravitation.
Once the metric tensor had been adopted as the central dynamical quantity, it came to embody the link between physical measurements and the structure of spacetime itself.

In modern treatments, the  tensor $\frak{g}$ is understood as a geometrical object, encoding the causal and metrical structure of spacetime. The invariant interval $ds^2$ had already been introduced by Minkowski in 1908, where he identified it with the proper time $s = \int ds$ measured along a particle’s worldline. As Minkowski emphasized in his {\it Raum und Zeit} lecture \cite{Minkowski}, $ds$ corresponds to the {\it Eigenzeit}, the proper time experienced by the particle. In Minkowski’s hands, however, $ds^2$ functioned primarily as an algebraic device within his four-dimensional reformulation of Special Relativity, which emphasized unification rather than geometry. It was Einstein who, a few years later, endowed the invariant with a genuinely geometrical meaning. In his 1916 review \cite{Einstein1916}, he wrote:
\begin{quote}
“{\it Borrowing from Gauss’s theory of surfaces, we have called ($ds$) the ‘linear element’}”, 
\end{quote}
embedding the interval within the language of Riemannian geometry. While Minkowski interpreted $ds$ as the infinitesimal proper time, Einstein progressively reinterpreted it as an element of the geometrical structure of spacetime—a shift that would later obscure its original dynamical meaning. In General Relativity, this interpretation requires care: $ds^2$ can be positive, negative, or zero, corresponding to timelike, spacelike, or null separations. Geometrically, this does not yield distances in the Riemannian sense, but rather classifies intervals and defines the light-cone structure that governs causality.

From this perspective, curvature characterizes how the relative motion of freely falling particles departs from uniform parallelism. Mathematically, this is expressed through the Riemann tensor, whose contraction yields the Ricci tensor appearing in the field equations. Physically, curvature manifests as tidal forces: nearby free-falling particles experience relative accelerations induced by the presence of matter and energy. In this sense, the “geometrization” of gravity should be understood less as an ontological claim about spacetime itself and more as a reformulation of the dynamics of matter and inertia in terms of the intrinsic structure of the manifold.

{Moreover, the hole argument \cite{Stachel2014,Iftime-Stachel,Earman-Norton,Weatherall} cautions against interpreting spacetime as a literal physical entity. Einstein realized in 1913–15 that if the field equations are generally covariant, then for a fixed matter distribution there exist infinitely many mathematically distinct solutions, related by active diffeomorphisms acting inside an empty ``hole.'' Assuming that the metric tensor represents a physical substance would imply an apparent non-uniqueness, undermining determinism in a Machian sense, where a given mass distribution must determine inertia. Einstein’s resolution was to abandon substantivalism and recognize that all diffeomorphically related solutions are physically equivalent. In this view, the spacetime manifold with its pseudo-Riemannian metric is not a material object but a mathematical structure encoding the relational dynamics between matter and inertia.

Crucially, this relational perspective has direct implications for experimental physics. Observable quantities must be diffeomorphism-invariant: measurements, observations, and predictions can only refer to relations between fields or events, rather than to the value of a field at a particular coordinate. For instance, in gravitational experiments one does not measure $\frak{g}(P)$ at a given event $P=(t,{\bf x})$, but rather the relative motion of test particles, the redshift between clocks, or the time delay of light signals. By identifying diffeomorphic solutions as physically equivalent, Einstein adopted a relational conception of spacetime via his notion of ``point coincidences,'' which resolves the hole argument and ensures that the operational content of measurements is consistent with the geometric formulation of gravity, where solutions of the field equations belong to equivalence classes of physically identical configurations.

}

This point becomes clearer when considering alternative but equivalent formulations of General Relativity. For instance, in {\it teleparallel gravity} \cite{Weitzenbock}, gravitation can be described not through curvature, but through torsion. In this framework, the Riemann tensor vanishes identically, while the gravitational interaction is encoded in a torsion tensor derived from a tetrad field. The field equations in the teleparallel approach are dynamically equivalent to Einstein’s equations, even though the underlying geometrical picture is entirely different: spacetime is flat, but endowed with torsion. This equivalence highlights that what matters physically is not whether gravity is represented as curvature or torsion, but that test particles and matter fields evolve according to the same dynamics. In other words, the geometrical representation of gravitation is ultimately a matter of convention rather than of substance.

{From a technical point of view, teleparallel gravity offers certain advantages,
such as a gauge-theoretic formulation and the possibility of defining a genuine
gravitational energy--momentum tensor, as well as a clearer separation between
inertial and gravitational effects. On the other hand, the theory requires the
introduction of tetrad fields, and issues related to local Lorentz invariance can
arise, particularly in modified teleparallel models. In its standard form, the
teleparallel equivalent of General Relativity is experimentally
indistinguishable from General Relativity, and any observable deviations would
only appear in extensions beyond the dynamically equivalent regime.

}

A similar caution was expressed by Einstein in his 1920 Leiden lecture \cite{Einstein1920}, where he introduced the notion of an ``ether'' in General Relativity. Unlike the mechanical ether of pre-relativistic physics, this was not a medium of particles or endowed with mechanical properties, but rather a way of saying that spacetime in GR possesses metrical and causal structure given by the metric tensor. Einstein explicitly warned that it would be misleading to imagine spacetime as a flexible substance that bends and curves like an elastic sheet. Instead,  ``ether'' in GR refers to the set of mathematical relations encoded in the line element $ds^2$—or, in Minkowski’s perspective, the infinitesimal proper time—that govern the motion of test particles and the propagation of fields. This underlines once more that the geometrical description is a formal framework, not a literal ontology of spacetime as a physical medium.

Interestingly, this interpretation brings Einstein’s mature view into close alignment with later field-theoretic perspectives such as that of Weinberg. In his \textit{Gravitation and Cosmology}  \cite{Weinberg}, Weinberg emphasized that the geometric formulation of General Relativity should not be regarded as an ontological statement about space and time, but merely as a convenient representation of a physical field whose observable effects—light deflection, time dilation, planetary motion—could equally well be attributed to curvature or to the dynamics of a tensor field in flat spacetime. Although Weinberg presented this as a heterodox view, it actually echoes Einstein’s own conclusions after the hole argument and the Leiden lecture: that the metric field does not describe a material geometry, but a relational state of a dynamical entity governing the behaviour of matter and inertia. Both perspectives converge on the idea that the “geometry” of spacetime is a manifestation of physical relations rather than an underlying substance—a viewpoint that dissolves the apparent opposition between the geometric and dynamical interpretations of gravitation.

{

}

In summary, while the prevailing interpretation of GR emphasizes its geometric character, it is important to recall that Einstein himself arrived at the theory by insisting on physical principles. Only later did he realize that the mathematical formalism of geometry provided the natural language in which those principles could be consistently expressed—a distinction often blurred in later interpretations of General Relativity.

\section{Rethinking Gravitation: Dynamics Without Geometry in the Weak-Field Limit}

Having clarified that Einstein’s geometrical language emerged as a formal representation of deeper physical principles, we may now, in the spirit of Weinberg’s field-theoretic approach \cite{Weinberg}, ask whether these same principles—particularly the Equivalence Principle and the Principle of Relativity—can be reformulated in a purely dynamical manner, without assigning a privileged role to differential geometry. The key question is whether gravitation must necessarily be identified with the geometry of a curved manifold or whether its observable effects can instead be understood as the manifestation of a physical field whose dynamics, within a locally Lorentz-invariant framework, express the equivalence between inertia and gravitation.

In this alternative view, geometry ceases to play an ontological role and emerges instead as an effective representation of the underlying relational dynamics between matter and inertia, encoded in the behavior of the metric or an equivalent tensorial field.

With this motivation, our goal is to incorporate gravitation into the framework of Special Relativity by replacing the Minkowski invariant with a modified one that includes the gravitational potential. The requirement is that the resulting equations of motion reproduce, as closely as possible, Newton’s second law, thereby making explicit the dynamical equivalence between gravitational and inertial forces \cite{HE2025}. 

To gain intuition about the appropriate form of this invariant, we begin by applying the Equivalence Principle to the case of a uniform gravitational potential, $\Phi = gz$, and examine how Newton’s second law can be recovered in the spirit of D’Alembert \cite{Dalembert}:
\begin{quote}
\begin{flushleft}
\textit{In a system, the internal inertial forces are equal and opposite to the forces that produce the acceleration}.

(D’Alembert, 1757)
\end{flushleft}
\end{quote}

Therefore, we can write
\begin{eqnarray}\label{inertia}
{\bf F}_{\rm i} + {\bf F}_{\rm g} = 0,
\end{eqnarray}
where the inertial force is ${\bf F}_{\rm i} = -m{\bf a}_{\rm p} \equiv - m(0,0,a_{\rm p})$ and the gravitational force is ${\bf F}_{\rm g} = -m(0,0,g)$. Here $a_{\rm p}$ denotes the acceleration of the body along the $OZ$ axis, which we will later identify with the spatial component of the four-acceleration in Special Relativity. In writing this balance we have also assumed the numerical equality of inertial and gravitational masses.

\begin{remark}
To properly understand D’Alembert’s Principle, it is useful to recall Newton’s concept of \textit{vis insita}, introduced in the \textit{Principia} \cite{Newton}:
\begin{quote}
\begin{flushleft}
\textit{``The vis insita, or innate force of matter, is a power of resistance by which every body, as far as it is able, continues in its present state, whether it be of rest or of moving uniformly forward in a straight line. This force is always proportional to the body whose force it is, and differs nothing from the inactivity of the mass, but in our way of conceiving it.''}
\end{flushleft}
\end{quote}

In this definition, Newton attributes to matter an intrinsic dynamical property—the \textit{vis insita}—by which each body resists any change in its state of motion. Although Newton describes this property as a “power of resistance,” it implicitly contains the idea of an internal reaction that manifests whenever an external agent attempts to accelerate the body.

D’Alembert made this implicit idea explicit. In his \textit{Traité de dynamique} (1743) \cite{Dalembert}, he identified the \textit{vis insita} with the inertial force, treating it as a real and active counterpart to the applied forces. He thus reformulated the laws of motion in the following terms:
\begin{quote}
\begin{flushleft}
\textit{``To reduce problems of dynamics to problems of statics, one may consider that the inertial forces $-m\mathbf{a}$ act on the body, opposing the applied forces.''}
\end{flushleft}
\end{quote}

In this formulation, the inertial force $-m\mathbf{a}$ is not a fictitious quantity but the dynamical expression of the body’s \textit{vis insita}. It represents the reaction of matter to acceleration, acting in opposition to the impressed forces so that, in D’Alembert’s view, motion results from a state of dynamical equilibrium:
\begin{equation}
    \sum \mathbf{F}_{\text{applied}} + (-m\mathbf{a}) = 0.
\end{equation}

D’Alembert’s insight thus transforms Newton’s second law into a principle of balance between real forces—external (gravitational, elastic, etc.) and internal (the inertial reaction). This reinterpretation, grounded in the identification of \textit{vis insita} with inertia, anticipates the modern form of the Equivalence Principle, where the inertial and gravitational forces exactly compensate each other in free fall.
\end{remark}



{In close analogy with General Relativity—where the motion of free particles is encoded in the variational principle $\delta \int ds = 0$—we adopt Minkowski’s original interpretation of the spacetime interval $ds$ as the infinitesimal measure of proper time \cite{Minkowski}. From this perspective, dynamics is not imposed externally through forces, but emerges from a variational structure intrinsic to spacetime itself. Building on Planck’s variational formulation of Special Relativity \cite{Planck}, we reinterpret Fermat’s principle for light propagation \cite{Fermat}—according to which light selects paths of extremal travel time—as a deeper organizing principle of relativistic motion. We then extend this variational viewpoint to massive bodies by postulating that the trajectory of a test particle in a gravitational field is the one that extremizes its proper time. In this sense, gravitation is not introduced as a force acting in spacetime, but rather as a modification of the relational structure that determines which worldlines are dynamically preferred.

}

Thus, considering the isotropic invariant:
\begin{eqnarray}\label{uniform}
    ds^2 = A(z) dt^2 - B(z) d{\bf x} \cdot d{\bf x},
\end{eqnarray}
and minimizing the proper time ${ \delta \int ds = 0}$ for free fall along the $OZ$ direction,
{which is equivalent to minimizing \cite{Landau}
\begin{eqnarray}
    \int L ds\qquad \mbox{with}\qquad
    L=A(z)\left(\frac{dt}{ds}\right)^2-B(z) \left|\frac{d{\bf x}}{ds}\right|^2,    \end{eqnarray}
together with $L=1$}
{yields, via the Euler-Lagrange equations, the corresponding dynamical equation is:}
\begin{align}
  m \frac{d^2 z}{ds^2} = - \frac{m}{2 A(z) B(z)} \left[
    \partial_z A + \partial_z (A B) \left( \frac{dz}{ds} \right)^2 \right].
\end{align}

To recover Newton’s second law, it is necessary to impose that the velocity-dependent term on the right-hand side vanish. This condition holds when
$
{A(z) B(z) = C},
$
where ${C}$ is a constant that can be taken as unity by rescaling coordinates. 

Thus, we obtain:
\begin{eqnarray}
  - m \frac{d^2 z}{ds^2} - \frac{m}{2} \partial_z A = 0,
\end{eqnarray}
and comparison with (\ref{inertia}) yields
\begin{eqnarray}\label{exact}
a_{\rm p} = \frac{d^2 z}{ds^2}, \qquad 
A(z) = b + 2 \Phi(z), \qquad B(z) = \frac{1}{b + 2 \Phi(z)},
\end{eqnarray}
where ${b}$ is a constant that must equal unity to recover the Minkowski invariant when the field vanishes.

We conclude, therefore, that for a uniform gravitational field described by ${\Phi(z) = gz}$, a freely falling object obeys Newton’s second law when the acceleration $a_{\rm p}$ represents the spatial component of the four-acceleration ($\frac{d^2z}{ds^2}=-g$).

\medskip

This observation connects directly with Einstein’s original 1907 formulation of the Equivalence Principle: a freely falling observer does not experience their own weight. This insight may have been influenced by Ernst Mach, who described the sensation of falling as the disappearance of the internal pressures that the body’s particles exert on one another under the action of weight \cite{Mach}:
\begin{quote}
\textit{When we jump or fall from a certain height, we experience a particular sensation that must arise from the cessation of the pressures that the particles of the body (from blood, etc.) exert on one another due to their weight}.
\end{quote}

From this standpoint, the sensation of weightlessness does not imply that gravitation ceases to act. Rather, as Mach’s description suggests, the gravitational and inertial forces exactly compensate each other during free fall. The body continues to be accelerated by gravity, but its internal stresses vanish because this acceleration is dynamically balanced by its own inertial reaction. In this interpretation, the Equivalence Principle does not assert the disappearance of gravity, but the exact compensation between two real and opposite forces—gravitational and inertial—acting on the same body. Free-fall thus represents a state of dynamical equilibrium, not the absence of gravitation.

Regarding the question “with respect to what is the body accelerated?”, it is important to emphasize that the inertial force is not measured with respect to an absolute space, as in Newtonian mechanics, but relative to the local gravitational environment. In weak-field systems such as the Solar System, the acceleration of a body can be naturally defined with respect to the local static gravitational field, which establishes the reference frame for the motion of test particles. More generally, following the relational perspective inspired by Mach, the inertial reaction of a body reflects its interaction with the surrounding matter: apparent acceleration is meaningful only in relation to other masses in the universe. In this framework, the inertial force $-m  d^2 z / ds^2$ represents a real dynamical reaction, and its exact compensation with the gravitational force during free fall explains the sensation of weightlessness without invoking absolute space. This provides a fully physical interpretation of Einstein’s first version of the Equivalence Principle \cite{Einstein1907}, in which acceleration, inertia, and gravity are dynamically interrelated rather than geometrically postulated.

This physical reading of the Equivalence Principle contrasts sharply with the modern geometric formulation that became standard after Pauli’s reinterpretation. Indeed, a crucial distinction must be made between Einstein’s original formulation of the Equivalence Principle (1907–1911) and its later local reinterpretation introduced by Pauli within the mature framework of General Relativity.

Einstein’s principle referred to a global physical equivalence between a uniformly accelerated frame and a uniform gravitational field. It expressed a genuine dynamical identity between inertial and gravitational forces: the inertial reaction $-m  d^2z/ds^2$ exactly compensates gravitational attraction when $m_{\rm i}=m_{\rm g}$. In this view, the state of weightlessness in free fall results from the balance of two real and opposite forces acting on the same body, rather than from the geometrical vanishing of gravity.

By contrast, for a general gravitational field, the version formulated by Pauli in 1921 introduced a fundamentally different concept. Pauli stated that \cite{Pauli1921}
\begin{quote}
\textit{
“For every infinitely small world region (i.e. a world region which is so small that the space- and time-variation of gravity can be neglected in it) there always exists a coordinate system … in which gravitation has no influence either on the motion of particles or any other physical process”}.
\end{quote}

This reformulation shifted the emphasis from dynamics to geometry.
The Equivalence Principle no longer expressed a physical identity between acceleration and gravitation, but a mathematical statement about the local properties of spacetime.
At each point, one can introduce coordinates in which the metric is locally Minkowskian and the Christoffel symbols vanish—precisely as one can introduce {\it geodesic normal coordinates} on a {Gaussian} curved surface so that, in an infinitesimal region, geodesics appear as straight lines and curvature becomes imperceptible.
In both cases, the apparent “absence” of curvature does not reflect a physical reality but a coordinate adjustment that hides it, at first order.
Pauli’s version thus transformed Einstein’s original principle—rooted in the dynamical interplay between inertia and gravitation—into a purely geometrical statement about the local flatness of spacetime.
The profound physical content of Einstein’s early conception, in which gravitation and inertia were two aspects of the same dynamical process, was  replaced by a kinematical property of the connection.

Even if this local statement is essential for the mathematical structure of General Relativity, it actually departs from Einstein’s 1907–1911 conception, which was global and physical rather than infinitesimal and geometric.
Einstein’s version described a genuine physical equivalence—between a homogeneous gravitational field and a uniformly accelerated system—valid over a finite region of spacetime.
Pauli’s formulation, by contrast, only guarantees that in an infinitesimal neighborhood of any point, the motion of free particles in a general gravitational field can be approximated as uniform and rectilinear, after an appropriate choice of coordinates.

The formulation developed here returns to Einstein’s original insight, interpreting the Equivalence Principle as a physical law that connects inertia and gravitation. From this standpoint, the gravitational modification of the invariant $ds^2$ is not postulated geometrically but dynamically derived from the balance of inertial and gravitational forces, preserving the physical content of Einstein’s early conception within a consistent relativistic framework.

\begin{remark}

The same dynamical procedure developed for the uniform gravitational field can be extended to the case of a spherically symmetric source. In this situation, the spacetime interval can be written in general static and isotropic form
\begin{equation}
ds^2 = A(r)dt^2 - B(r)dr^2 - r^2 d\Omega^2,
\end{equation}
where $d\Omega^2 = d\theta^2 + \sin^2 \theta d\phi^2$.

Applying the same variational principle—that free trajectories extremize proper time—to this configuration, and requiring that the resulting equation of motion reproduces Newton’s law in the non-relativistic limit, one obtains
\begin{eqnarray}\label{Droste}
ds^2 = \left(1-\frac{2MG}{r}\right) dt^2 -
\left(1-\frac{2MG}{r}\right)^{-1} dr^2 - r^2 d\Omega^2,
\end{eqnarray}
which exactly coincides with the result obtained by Johannes Droste \cite{Droste} through the direct integration of Einstein’s vacuum field equations.

For $r>2MG$, this invariant is physically equivalent to the one obtained independently by Karl Schwarzschild \cite{Schwarzschild}:
\begin{eqnarray}\label{Schwarzschild}
    ds^2 = \left(1-\frac{2MG}{f(r)}\right) dt^2 -
    \left(1-\frac{2MG}{f(r)}\right)^{-1} (f'(r))^2 dr^2 - f^2(r) d\Omega^2,
\end{eqnarray}
{where
$$
f(r) = \bigl(r^3 + (2MG)^3\bigr)^{1/3}.
$$

The equivalence between (\ref{Droste}) and (\ref{Schwarzschild}) follows from a
radial diffeomorphism acting on the metric field rather than from a passive
change of coordinates. Concretely, one considers an active diffeomorphism
$\varphi$ acting on the radial coordinate according to
$$
r \mapsto f(r)=\bigl(r^3+(2MG)^3\bigr)^{1/3},
$$
and defines a new metric $\mathfrak g'=\varphi^\ast\mathfrak g$ on the same
coordinate domain. Under this pullback, the angular and radial terms transform as
$r^2 d\Omega^2 \to f^2(r)d\Omega^2$ and $dr^2 \to (f'(r))^2dr^2$, yielding exactly
the form (\ref{Schwarzschild}).

This equivalence reflects the invariance of Einstein’s field equations under
active diffeomorphisms \cite{Macdonald}. In particular, the coordinate singularity
at $r=0$ in Schwarzschild’s original solution corresponds to $r=2MG$ in Droste’s
radial coordinate.

}

Schwarzschild actually found a one-parameter family of solutions, all physically equivalent, of which Droste’s solution is a particular case. He selected one singular at $r=0$, only. At the time (1916), the concept of invariance under active diffeomorphisms was not fully established, which led to confusion regarding the interpretation of coordinates \cite{Misner}.

A subtle but essential point concerns the relation between the Schwarzschild and Droste forms of the static, spherically symmetric solution. 
The standard view holds that these correspond to distinct coordinate systems related by a change of variables, but a more consistent interpretation regards them as physically equivalent  fundamental forms $\mathfrak{g}$ defined in the same coordinate domain, connected by an \emph{active} diffeomorphism. 
This perspective directly reflects Einstein’s ``hole argument'': different $\mathfrak{g}$ defined over the same manifold may represent the same physical situation, provided they are related by an active diffeomorphism. 

In this sense, the Schwarzschild and Droste expressions are mathematically distinct but physically equivalent representations of the same gravitational field. 
What changes is not the physical content of the field but the \emph{interpretation of the coordinates} themselves, whose operational meaning is determined by the fundamental invariant. 
This viewpoint fits naturally with the dynamical interpretation of the Equivalence Principle developed here: the gravitational field defines the local structure of inertia and thereby the physical meaning of the coordinates. 
Spacetime points, therefore, acquire physical significance only through the coefficients of the invariant, not independently of it.

Finally, as a historical note, it was David Hilbert \cite{Hilbert} who first named the metric (\ref{Droste}) the ``Schwarzschild solution.'' Today, Droste’s form is universally adopted in textbooks, even though the name still honors Schwarzschild. This reflects historical convention rather than scientific priority: the metric commonly called ``Schwarzschild solution'' corresponds in fact to Droste’s invariant.
\end{remark}

\subsection{Static Fields}

{Building on the previous result for a uniform gravitational field, we now consider a conformastatic spacetime, that is, a static spacetime whose metric is conformally related to the flat Minkowski metric. In such spacetimes, both the temporal and spatial parts of the line element are rescaled by position-dependent conformal factors, reflecting the gravitational modification of proper time and spatial measurements while preserving spatial isotropy. 

Accordingly, for a static, pressureless configuration we adopt the following conformastatic invariant \cite{Synge}:}
\begin{eqnarray}\label{conformastat}
   ds^2=(1+2\Phi({\bf x}))\,dt^2-(1+2\Phi({\bf x}))^{-1} d{\bf x}\cdot d{\bf x}.
\end{eqnarray}

{For this conformastatic invariant,}
the minimization of the proper time, 
{is equivalent to minimizing 
\begin{eqnarray}
    \int L ds\qquad \mbox{with}\qquad
    L=(1+2\Phi({\bf x}))\dot{t}^2-(1+2\Phi({\bf x}))^{-1} |\dot{\bf x}|^2,    \end{eqnarray}
together with $L=1$.}
{The Euler-Lagrange equations yield:}
\begin{eqnarray}\label{geodesicequation}
   \ddot{\bf x}
    =-\nabla\Phi
    -2
    \frac{|\dot{\bf x}|^2}{1+2\Phi}\nabla^{\perp}\Phi ,
\end{eqnarray}
where the dot denotes the derivative with respect to the proper time and 
\begin{eqnarray}
\nabla^{\perp}\Phi=
\nabla\Phi-\frac{\dot{\bf x}\cdot \nabla\Phi}{|\dot{\bf x}|^2}\dot{\bf x}
\end{eqnarray}
is the orthogonal projection of $\nabla\Phi$ onto the subspace orthogonal to $\dot{\bf x}$.

\

When the velocity is parallel to the gradient of $\Phi$, {one has 
$\nabla^{\perp}\Phi=
\nabla\Phi-\frac{\nabla\Phi\cdot \nabla\Phi}{|\nabla \Phi|^2}\nabla\Phi=0$, 
and thus the exact Newtonian equation is recovered:} 
\begin{eqnarray}
   \ddot{\bf x}=-\nabla\Phi ,
\end{eqnarray}
which reflects the { Weak Principle of Equivalence}.  
In this framework, the inertial force ---the Newtonian {\it vis insita} \cite{Newton}--- acting on a particle of inertial mass $m_{\rm i}$ is identified with the spatial component of the four-force $-m_{\rm i}\ddot{\bf x}$.  
Denoting by $m_{\rm g}$ the gravitational mass, the dynamical equation becomes a balance of forces:
\begin{eqnarray}
   {\bf F}_{\rm i}+{\bf F}_{\rm g}=0 
   \quad \Longrightarrow \quad 
   -m_{\rm i}\ddot{\bf x}-m_{\rm g}\nabla\Phi=0,
\end{eqnarray}
with $m_i=m_g$.

\

As an example, consider a freely falling body near the Earth’s surface moving along a purely radial trajectory,  
${\bf x}(t)=x(t){\bf x}$.  
In this case, the exact equation of motion reduces to
\begin{eqnarray}
   \ddot{x}=-\frac{MG}{x^2}\cong 9.8~{\rm m/s^2},
\end{eqnarray}
in  {International System of Units}.  
This illustrates the equivalence of gravitational and inertial effects:  
the inertial force, corresponding to the four-force, exactly cancels the gravitational one.  
Thus, the body does not feel its own weight—a manifestation of Einstein’s early formulation of the Equivalence Principle.

\

Moreover, the conformastatic invariant naturally implies the conservation of energy.  
Indeed, taking the inner product of equation \eqref{geodesicequation}, with the momentum ${m\dot{\bf x}}$, yields
\begin{eqnarray}
   \frac{m|\dot{\bf x}|^2}{2}+m\Phi({\bf x})=E,
\end{eqnarray}
{because $
\dot{\bf x}\cdot\nabla^{\perp}\Phi=
\dot{\bf x}\cdot\nabla\Phi-\frac{\dot{\bf x}\cdot \nabla\Phi}{|\dot{\bf x}|^2}\dot{\bf x}\cdot\dot{\bf x}=\dot{\bf x}\cdot\nabla\Phi-\dot{\bf x}\cdot\nabla\Phi=
0.
$

}

\

Another compelling feature of this metric is that it predicts the same deflection of light as the Droste solution, namely \cite{Weinberg}
\begin{eqnarray}
   \delta\phi=\frac{4MG}{b},
\end{eqnarray}
where $b$ denotes the impact parameter.

\

The key point can already be seen in Newtonian gravity, where the second Newton’s law
\begin{eqnarray}
   \frac{d^2{\bf x}}{dt^2}=-\nabla\Phi ,
\end{eqnarray}
after identifying gravitational and inertial mass, can be written as
\begin{eqnarray}
   -m\frac{d^2{\bf x}}{dt^2}-m\nabla\Phi=0
   \quad \Longrightarrow \quad 
   {\bf F}_{\rm i}+{\bf F}_{\rm g}=0 .
\end{eqnarray}
That is, inertial and gravitational forces compensate, so the body does not feel its weight.

\

The same occurs in a metric theory, including relativity, where the inertial force must be replaced by the four-force—the real force acting on the body.  
Accordingly, the geodesic equation for a conformastatic metric takes the form
\begin{eqnarray}
   {\bf F}_{\rm i}+{\bf F}_{\rm g}+{\bf F}_{\rm tidal}=0,
\end{eqnarray}
where 
\begin{eqnarray}
   {\bf F}_{\rm tidal}=-2m\frac{|\dot{\bf x}|^2}{1+2\Phi}\nabla^{\perp}\Phi ,
\end{eqnarray}
{can be interpreted as the tidal force.}

Thus, when the velocity is parallel to the gradient of $\Phi$, one recovers exactly Newton’s law.

\subsection{Moving masses}

We now study the gravitational effect produced in the invariant $ds^2$ when the configuration is not static, that is, when the sources are in motion.  
Following Kenneth Nordtvedt \cite{Kenneth}, let us consider another frame $K'$ moving with three-dimensional velocity $-{\bf v}'$ with respect to the conformastatic frame $K$.  
Equivalently, $K$ moves with velocity 
\begin{eqnarray}
   {\bf v}'=\frac{d{\bf x}'}{dt'}
\end{eqnarray}
with respect to $K'$.

\

Let $\gamma$ be the Lorentz factor
\begin{eqnarray}
   \gamma=\frac{1}{\sqrt{1-|{\bf v}'|^2}} .
\end{eqnarray}
Then, the infinitesimal Lorentz transformation reads
\begin{eqnarray}
   d{\bf x}=\gamma\left[
    \frac{1}{\gamma}d{\bf x}'+\frac{\gamma}{1+\gamma}({\bf v}'\cdot d{\bf x}'){\bf v}'-{\bf v}'dt'
    \right],\qquad 
    dt=\gamma(dt'-{\bf v}'\cdot d{\bf x}').
\end{eqnarray}

{The key point is that, given two constants $a$ and $b$, the quantity $a\,dt^2 + b\, d{\bf x}\cdot d{\bf x}$ transforms into
\begin{eqnarray}
\gamma^2(a+b|{\bf v}'|^2)\, dt'^2
-2\gamma^2(a+b)({\bf v}'\cdot d{\bf x}')\, dt'
+ b\, d{\bf x}'\cdot d{\bf x}'
+ \gamma^2(a+b)({\bf v}'\cdot d{\bf x}')^2.
\end{eqnarray}

Then, removing the primes, the invariant in $K'$ can be written, after linearizing with respect to both the potential and the velocity, as
\begin{eqnarray}\label{LORENTZ}
ds^2 = (1+2\Phi)\, dt^2 - 8\, {\bf N}\cdot d{\bf x}\, dt - (1-2\Phi)\, d{\bf x}\cdot d{\bf x},
\end{eqnarray}
where, to incorporate retardation effects while remaining compatible with special relativity, the Newtonian potential for pressureless matter—following Abraham’s theory \cite{Abraham} (see also \cite{Nordstrom})—is given by
\begin{eqnarray}
\Phi(t,{\bf x}) = -G\int \frac{\rho(t_{\rm r},\bar{\bf x})}{|{\bf x}-\bar{\bf x}|}\, d\bar{V}
\qquad \Longrightarrow \qquad
\Box\Phi = -4\pi G \rho,
\end{eqnarray}
being $t_{\rm r}= t - |{\bf x}-\bar{\bf x}|$ the retarded time, and $\Box = \partial_{t}^2 - \Delta$ the d'Alembert operator.

In parallel, by analogy with the relativistic formulation of the Biot--Savart law, the gravitational vector ${\bf N}$ satisfies
\begin{eqnarray}
{\bf N}(t,{\bf x}) = -G\int \frac{\rho(t_{\rm r},\bar{\bf x})\, {\bf v}(t_{\rm r},\bar{\bf x})}{|{\bf x}-\bar{\bf x}|}\, d\bar{V}
\qquad \Longrightarrow \qquad
\Box{\bf N} = -4\pi G \rho\, {\bf v},
\end{eqnarray}
where the source term corresponds to the mass-current density.

To obtain the explicit form of the gravitational potentials generated by a continuous distribution of matter we have to take into account that the contributions of each infinitesimal element are additive: each element acts independently in shaping the invariant. Since different elements can possess different velocities, the appropriate Lorentz transformation must be applied individually to each, according to its local state of motion. The complete expression is then obtained by combining these contributions, formally integrating over the entire distribution.

Specifically, one can divide the matter content into $N$ small cells, each with density $\rho_k$ and volume $\Delta\bar{V}_k$. The invariant can then be approximated as
\begin{eqnarray}
ds^2 \simeq \sum_{k=1}^N
\Bigg[
\Big(\frac{1}{N} +2\Phi_k \Big) dt^2
-
\Big( \frac{1}{N} - 2\Phi_k \Big) d{\bf x}\cdot d{\bf x}
\Bigg],
\end{eqnarray}
where we have introduced the notation
$$\Phi_k({\bf x}) \equiv -G\, \frac{\rho_k \Delta \bar{V}_k}{|{\bf x}-\bar{\bf x}_k|}.$$

}



\

Applying the appropriate Lorentz transformation to each of these local contributions and then taking the continuum limit precisely reproduces  the metric \eqref{LORENTZ}. This shows that the dynamical reinterpretation together with the conservation of the stress-tensor (\ref{stress-tensor}) is fully consistent with General Relativity in the weak-field regime.

\

{When all terms in the velocity are retained and retardation effects are taken into account, the line element acquires the form
\begin{eqnarray}\label{Invariant}
ds^2 &=&
(1 - 2\Phi + 4\Upsilon)\, dt^2
- 8\, \bar{\mathbf{N}} \cdot d\mathbf{x}\, dt
- (1 - 2\Phi)\, d\mathbf{x} \cdot d\mathbf{x}
+ 4\, \frak{t}(d{\bf x}, d{\bf x}),
\end{eqnarray}
where the three remaining potentials, $\Upsilon$, $\bar{\bf N}$ and $\mathfrak{t}$, obey
\begin{eqnarray}
\Box \Upsilon &=& -\frac{4\pi G \rho}{1- |\mathbf{v}|^2}, \\
\Box \bar{\bf N} &=& -\frac{4\pi G \rho}{1- |\mathbf{v}|^2}\, {\bf v}, \\
\Box \mathfrak{t} &=& -\frac{4\pi G \rho}{1-|\mathbf{v}|^2}\,
{\bf v}^{\flat} \otimes {\bf v}^{\flat},
\end{eqnarray}
where the ``flat'' operator is used to obtain the dual one-form associated with the vector field ${\bf v}$.

Therefore, expression~\eqref{Invariant} coincides with the result obtained in linearized General Relativity in the harmonic gauge,
when the stress-tensor is considered in Minkowski space, and thus, 
 the four-velocity is given by
\begin{eqnarray}
{\bf u}=\frac{1}{\sqrt{1-|{\bf v}|^2}}(1, {\bf v}).
\end{eqnarray}

This correspondence confirms that the present formulation reproduces, in the weak-field limit, the same physical content as linearized General Relativity in the harmonic gauge, while avoiding the explicit introduction of curvature. Hence, the geometric structure of spacetime emerges here as an effective description of the underlying dynamical relations between matter and inertia.

It is worth noting that, in the static case, the invariant (\ref{conformastat}) already satisfies the harmonic gauge, and since this gauge is preserved under Lorentz boosts, the boosted (i.e., moving-source) form of the invariant necessarily satisfies the same condition. This explains why the expressions obtained for moving masses coincide with those of linearized GR formulated in the harmonic gauge.

}

{

}

\begin{remark}
Neglecting the tensor potentials $\Upsilon$ and $\mathfrak{h}$ in (\ref{Invariant}), and following the analogy with electrodynamics, one may introduce the \textit{gravitational four-potential}
\[
    {\bf A} \equiv (\Phi, \mathbf{N})
\]
and the \textit{four-current}
\[
    {\bf J} \equiv (\rho, \rho\, \mathbf{v}),
\]
so that the field equations for the potentials take the compact and manifestly covariant form
\begin{equation}\label{XXX}
        \Box {\bf A} = -4\pi G\, {\bf J}.
\end{equation}

If the relativistic continuity equation $\mathrm{div}(\rho{\bf u}) = 0$ is approximated by its classical form,
\begin{eqnarray}
\partial_t \rho + \nabla \cdot (\rho \mathbf{v}) = 0 
\qquad \Longleftrightarrow \qquad 
\mathrm{div}({\bf J}) = 0,
\end{eqnarray}
this condition is equivalent to adopting the so-called \textit{Lorenz gauge},
\begin{eqnarray}
 \mathrm{div}({\bf A}) = 0 
 \qquad \Longleftrightarrow \qquad 
 \partial_t \Phi + \nabla \cdot \mathbf{N} = 0,
\end{eqnarray}
originally introduced in 1867 by the Danish physicist Ludvig Valentin Lorenz.  
It is worth recalling that the name is frequently misspelled as “Lorentz gauge,” owing to its similarity with Hendrik Antoon Lorentz, whose contributions concern classical electrodynamics and its relativistic transformations.

Moreover, one can define the \textit{gravitoelectromagnetic fields}:
\begin{eqnarray}
    {\bf E} \equiv -\nabla \Phi - \partial_t {\bf N}, 
    \qquad 
    {\bf H} \equiv \nabla \wedge {\bf N},
\end{eqnarray}
which, by virtue of their definitions and of Equation~(\ref{XXX}), satisfy the system of equations:
\begin{eqnarray}
    \nabla \cdot {\bf E} &=& -4\pi G \rho, 
    \qquad 
    \nabla \cdot {\bf H} = 0, 
    \nonumber\\
    \nabla \wedge {\bf E} + \partial_t {\bf H} &=& 0, 
    \qquad 
    \nabla \wedge {\bf H} - \partial_t {\bf E} = -4\pi G \rho\, {\bf v}.
\end{eqnarray}
These are the \textit{gravitoelectromagnetic equations}, the direct gravitational analogue of Maxwell’s equations in the weak-field limit.  
They make explicit the close formal correspondence between the linearized gravitational field and electromagnetism, confirming that the gravitational interaction can be expressed in a field-theoretic language without invoking curvature explicitly.
\end{remark}

Therefore, this procedure ensures that the resulting invariant correctly accounts for the collective influence of all moving sources, yielding the same result as linearized GR in the harmonic gauge, without the need to invoke the full geometric formalism of General Relativity. In the spirit of Fock \cite{Fock}, one may interpret this result as highlighting the physical significance of harmonic coordinates, which he argued to provide the most natural description of the gravitational field. Within this perspective, the harmonic gauge is not merely a convenient mathematical choice, but it reflects a preferred set of coordinates in which the dynamical effects of gravity—encoded in the invariant—are most directly and transparently represented. This reinforces the view that the physical content of General Relativity can, at least in the weak-field regime, be captured entirely through a properly defined field-theoretic invariant, in harmony with the Equivalence Principle and Lorentz invariance, without invoking full Riemannian geometry.

{

{

{

}

}

{
\subsection{Matter supporting pressure}
\label{sec-2}

{

To extend the previous analysis beyond pressureless matter,
we now turn to matter distributions capable of supporting internal pressure.
A convenient starting point is the Newtonian description of a homogeneous spherical body that may expand or contract within ordinary Euclidean space.
This simple setting already incorporates, in a consistent way, the relation between density, pressure, and the gravitational potential.

Let the sphere have radius 
$a(t)R$
 and homogeneous mass density 
$\sigma_0(t)$.
The total mass contained within it is therefore
\begin{equation}
M = \frac{4\pi}{3}\,\sigma_0(t) a^3(t)R^3.
\end{equation}

Choosing the center of the ball as the origin, we denote the position of a generic point $P$ by $a(t){\bf x}$,
so that its physical distance from the center is
$r \equiv a(t)|{\bf x}|$.

\medskip

By the shell theorem \cite{Newton} (see also p.~61 of \cite{Ryden}), only the mass enclosed within the radius 
$r$ contributes to the gravitational force.
A test particle momentarily at rest at 
$P$ thus feels a radial force
\begin{equation}
{\bf F} = -\frac{4\pi G m}{3}\,\sigma_0(t)\,a(t){\bf x},
\end{equation}
where $m$ is its mass.

Applying Newton’s second law and using the radial nature of the force, we obtain
\begin{equation}
m\,\frac{d^2 r}{dt^2} 
= -\frac{4\pi G m}{3}\,\sigma_0\,r
\qquad\Longrightarrow\qquad
\frac{d^2 r}{dt^2} 
= -\frac{G\,M(r)}{r^2},
\end{equation}
where $M(r)=\frac{4\pi}{3}r^3 \sigma_0$ is the mass enclosed within the sphere of radius $r$.

\medskip

This equation may also be obtained from a variational principle.  
Consider the Newtonian Lagrangian
\begin{equation}
L = \frac{1}{2}\left(\frac{dr}{dt}\right)^2 
+ \frac{4\pi G}{3}\, r^2 \sigma_0
\qquad\Longleftrightarrow\qquad
L = \frac{1}{2}\left(\frac{dr}{dt}\right)^2 - V(r),
\end{equation}
with potential
\begin{equation}
V(r) = -\frac{G M(r)}{r}.
\end{equation}

The Euler--Lagrange equation yields
\begin{equation}
\frac{d^2 r}{dt^2} = \frac{4\pi G}{3}\,\frac{d}{dr}(r^2 \sigma_0).
\end{equation}

Assuming mass conservation inside a ball of radius $r$ implies
\begin{equation}
\frac{d M(r)}{dt}=0\quad \Longleftrightarrow\quad
\frac{d}{dt}\!\left(\frac{4\pi}{3} r^3 \sigma_0\right) = 0
\quad\Longleftrightarrow\quad
\frac{dr}{dt}\frac{d}{dr}(r^3\sigma_0)=0
\quad\Longleftrightarrow\quad
\frac{d}{dr}(r^3\sigma_0)=0,
\end{equation}
and therefore
\begin{equation}
\frac{d}{dr}(r^2 \sigma_0) = - r \sigma_0.
\end{equation}
Substituting into the Euler--Lagrange equation gives
\begin{equation}
\frac{d^2 r}{dt^2} = -\frac{4\pi G}{3}\, \sigma_0\, r.
\end{equation}

\medskip

To incorporate pressure and obtain a general fluid description, the internal energy and its thermodynamic relation with volume must be included. This is achieved by replacing the mass density with the homogeneous total energy density $\rho_0$, leading to the modified Newtonian Lagrangian~\cite{Harko}
\begin{equation}
\bar{L} = \frac{1}{2}\left(\frac{dr}{dt}\right)^2 
+ \frac{4\pi G}{3}\, r^2 \rho_0.
\end{equation}

Using the Euler--Lagrange equation together with the first law of thermodynamics applied to the spherical volume of radius $r$,
\begin{equation}
d\!\left(\frac{4\pi}{3}r^3\rho_0\right)
=-p_0\, d\!\left(\frac{4\pi}{3}r^3\right)
\quad\Longleftrightarrow\quad
d(r^3 \rho_0) = - p_0\, d(r^3)
\quad\Longleftrightarrow\quad
rd(r^2 \rho_0)+r^2\rho_0dr = - 3p_0r^2\, dr,
\end{equation}
we obtain
\begin{equation}
\frac{d}{dr}(r^2 \rho_0)=-(\rho_0+3p_0)\,r.
\end{equation}

The equation of motion then becomes the well-known  second Friedmann equation~\cite{Friedmann}
\begin{equation}
\frac{d^2 r}{dt^2} 
= -\frac{4\pi G}{3}\,(\rho_0 + 3p_0)\, r
\qquad\Longrightarrow\qquad
\frac{d^2 r}{dt^2}
= -\frac{G\,M_{\rm g}(r)}{r^2},
\end{equation}
where
\begin{equation}
M_{\rm g}(r) 
= \frac{4\pi}{3} r^3 (\rho_0 + 3p_0)
\end{equation}
is the \textit{active gravitational mass}, introduced in Newtonian cosmology in~\cite{McCrea1951,Callan} and in the context of GR in~\cite{Tolman}.  
Thus the combination $\rho+3p$ acts as the \textit{active gravitational mass density}.

}

In particular, when matter supports pressure,  the Newtonian potential, namely  $\Phi_1$,   satisfies
\begin{equation}
\Box \Phi_1 = -4\pi G\,(\rho + 3p).
\end{equation}

\medskip

Now consider the stress--energy tensor
\begin{equation}
\mathfrak{T} = (\rho + p)\, {\bf u}^{\flat} \otimes {\bf u}^{\flat} - p\, \mathfrak{g},
\end{equation}
where ${\bf u}^{\flat}$
denotes the one-form obtained by lowering the index of the four-velocity
${\bf u}$ with the first fundamental form. It is defined by
\begin{eqnarray}  {\bf u}^{\flat}({\bf w})=\frak{g}({\bf u},{\bf w})\qquad
\mbox{for all four-vector }
{\bf w}.
\end{eqnarray}

The conservation law $\mathrm{div}(\mathfrak{T})=0$ in the Minkowski background $ds^2 = dt^2 - d{\bf x}\cdot d{\bf x}$ yields
\begin{equation}
\frac{d\rho}{ds} = -(\rho + p)\,\mathrm{div}({\bf u}),
\qquad
(\rho + p)\,\frac{d{\bf u}}{ds} = \mathrm{grad}(p) - \frac{dp}{ds}\,{\bf u},
\end{equation}
where $(\rho + p)\,d{\bf u}/ds$ is the relativistic four-force. Thus $\rho_{\rm in}=\rho+p$ is the \textit{effective inertial mass density} in the sense of Weinberg and Landau--Lifshitz~\cite{Weinberg, Landau}.

\medskip

These considerations motivate, in the static case, introducing a linearized invariant
\begin{equation}
ds^2 = (1 + 2\Phi_1)\, dt^2 - (1 - 2\Phi_2)\, d{\bf x}\cdot d{\bf x},
\end{equation}
with potentials obeying
\begin{equation}
\Box \Phi_1 = -4\pi G (\rho + 3p),
\qquad
\Box \Phi_2 = -4\pi G (\rho + b p),
\end{equation}
where $b$ is a dimensionless constant to be fixed physically.

After the Lorentz transformation described earlier, one obtains to linear  order with respect to both the potential
and the velocity,
\begin{equation}
ds^2 = (1 + 2\Phi_1)\, dt^2
- 8\, \tilde{\bf N}\cdot d{\bf x}\, dt
- (1 - 2\Phi_2)\, d{\bf x}\cdot d{\bf x},
\end{equation}
with the gravitational vector potential satisfying
\begin{equation}
\Box \tilde{\bf N} =
-4\pi G\!\left(\rho + \frac{1}{2}(3 + b)\, p\right){\bf v}.
\end{equation}

Since the vector potential is sourced by the current density $\sim(\rho+p){\bf v}$, consistency requires $b=-1$. Thus
\[
\Box \Phi_2 = -4\pi G (\rho - p),
\qquad
\Box \tilde{\bf N} = -4\pi G (\rho+p)\,{\bf v}.
\]

\medskip

Including all velocity-dependent terms, the invariant becomes
\begin{equation}\label{Invariant-full}
ds^2 = (1 - 2\Phi_2 + 4\hat{\Upsilon})\, dt^2
- 8\,\hat{\bf N}\cdot d{\bf x}\, dt
- (1 - 2\Phi_2)\, d{\bf x}\cdot d{\bf x}
+ 4\,\hat{\mathfrak{t}}(d{\bf x}, d{\bf x}),
\end{equation}
where
\begin{equation}
\Box \hat{\Upsilon}
= -\frac{4\pi G(\rho+p)}{1 - |{\bf v}|^2},\qquad
\Box \hat{\bf N}
= -\frac{4\pi G(\rho+p)}{1 - |{\bf v}|^2}\,{\bf v},\qquad
\Box \hat{\mathfrak{t}}
= -\frac{4\pi G(\rho+p)}{1 - |{\bf v}|^2}\,{\bf v}^{\flat}\!\otimes{\bf v}^{\flat}.
\end{equation}

This invariant reproduces exactly the weak-field expression derived in linearized General Relativity when the harmonic gauge is imposed. Equivalently, it coincides with the invariant governed by Einstein’s linear equation \cite{Einstein1955}:
\begin{eqnarray}\label{linearized}
    \Box \frak{h}=-16\pi G\left(\frak{T}-\frac{1}{2}\frak{\eta}T
    \right)\quad \mbox{with}\quad \frak{g}=\eta+\frak{h}.
\end{eqnarray}

The potentials $\hat{\Upsilon}$, $\hat{\bf N}$, and $\hat{\mathfrak{t}}$ are sourced by the inertial mass density $\rho+p$. The interpretation of $\Phi_2$ is subtler. Noting that
\[
\rho - p = 2(\rho + p) - (\rho + 3p),
\]
we introduce potentials $\hat{\Phi}_1$ and $\hat{\Upsilon}_1$, defined by
\[
\Box \hat{\Upsilon}_1 = -\frac{4\pi G (\rho+p)}{1 - |{\bf v}|^2}\,|{\bf v}|^2,
\qquad
\Box \hat{\Phi}_1 = -4\pi G (\rho+p),
\]
and obtain
\begin{equation}
\Phi_2 = 2\hat{\Phi}_1 - \Phi_1.
\end{equation}

In terms of these potentials, the invariant becomes
\begin{equation}\label{Invariant-split}
ds^2 = (1 + 2(\Phi_1 + 2\hat{\Upsilon}_1))\, dt^2
- 8\,\hat{\bf N}\cdot d{\bf x}\, dt
- (1 + 2(\Phi_1 - 2\hat{\Phi}_1))\, d{\bf x}\cdot d{\bf x}
+ 4\,\hat{\mathfrak{t}}(d{\bf x}, d{\bf x}).
\end{equation}

\medskip

Once the linearized form of the invariant $ds^2$ is written in terms of these gravitational potentials, the dynamics of continuous media follow from the conservation equation $\mathrm{div}(\mathfrak{T})=0$, supplemented by an Equation of State $p=p(\rho)$. This yields the relativistic continuity and Euler equations~\cite{Comer}:
\begin{equation}
\frac{d \rho}{ds} = -(\rho+p)\, \mathrm{div}(\mathbf{u}), 
\qquad
(\rho+p)\, \frac{D \mathbf{u}}{ds} 
= \mathrm{grad}(p) - \frac{dp}{ds}\, \mathbf{u}.
\end{equation}

}

 \section{ The Ricci tensor in harmonic gauge and the field equations}\label{Ricci}
In this section we present a compact, matrix-based formulation of the Ricci tensor in the harmonic gauge (also known as the de Donder/Lorenz gauge). This formulation avoids a large number of indices and expresses the nonlinear contributions to the Ricci tensor through matrix products, traces, and symmetrizations.

We will use the following matrix notation:
\begin{itemize}
\item $\frak{g}$ the matrix with components $g_{\mu\nu}$ (the first fundamental form),
\item $\frak{g}^{-1}$ its inverse,
\item $\Gamma^\alpha$ the matrix with components $(\Gamma^\alpha)_{\mu\nu}=\Gamma^{\alpha}_{\mu\nu}$ (Christoffel symbols),
\item $
(\partial_\alpha \frak{g})_{\mu\nu}=\partial_{\alpha} g_{\mu\nu}$,
\item $(\partial \frak{g}_{\alpha})_{\mu\nu}=\partial_\mu g_{\alpha\nu}$,
\item $\operatorname{Sim}[A]=\tfrac12 (A+A^T)$,
\item $\operatorname{Tr}$ the trace.
\end{itemize}

Let us now introduce the harmonic gauge (also called the De Donder or Lorenz gauge). To do so, we must start from the Laplace–Beltrami operator associated with
$\frak{g}$:
\begin{eqnarray}
\Box_{\frak{g}}\phi = \operatorname{div}(\operatorname{grad}\phi)=\frac{1}{\sqrt{-g}}
\partial_{\mu}\left(\sqrt{-g}g^{\mu\nu}\partial_{\nu}\phi \right),
\end{eqnarray}
where $g=\mbox{Det}(\frak{g})$ is the determinant of the first fundamental form.

We then consider the so-called harmonic coordinates defined by
\begin{eqnarray}
\Box_{\frak{g}} x^{\mu}=0, \qquad \mu=0,1,2,3.
\end{eqnarray}

In terms of the Christoffel symbols, this condition is equivalent to the harmonic gauge:
\begin{eqnarray}
\operatorname{Tr}( \frak{g}^{-1} \Gamma^{\alpha} )=0
\quad\Longleftrightarrow\quad
\mbox{Tr}\left(\frak{g}^{-1}\left(  \partial\frak{g}_{\alpha}-\frac{1}{2}\partial_{\alpha}\frak{g}     \right)
\right)=0
\quad\Longleftrightarrow\quad \partial_{\mu}(\sqrt{-g}g^{\mu\alpha})=0
, 
\quad \alpha=0,1,2,3,
\end{eqnarray}
where we have used that $\frak{g}$
is invertible. 

After a somewhat tedious calculation, one finds that in the harmonic gauge the Ricci tensor takes the form:
\begin{eqnarray}\label{Riccitensor}
R_{\mu\nu}
=
-\frac{1}{2}\,\Box_{\frak{g}} g_{\mu\nu}
-\operatorname{Tr}\!\left( \operatorname{Sim}[\partial_{\mu}\frak{g}^{-1}\,\partial \frak{g}_{\nu}] \right)
-\frac{1}{4}\operatorname{Tr}\!\left(
\frak{g}^{-1}\partial_{\mu}\frak{g}\,\frak{g}^{-1}\partial_{\nu}\frak{g}
\right).
\end{eqnarray}

This expression contains the linear part (the Laplace–Beltrami operator applied to
 $g_{\mu\nu}$)
 and all the quadratic terms in derivatives of 
  $\frak{g}$ 
  that account for the nonlinearity of the tensor.  
 
\vspace{0.2cm}

When the first fundamental form is a perturbation of that of Minkowski,
$\frak{g}=\eta+\frak{h}$, 
in the linear approximation one obtains:
\begin{eqnarray}
\frak{Ric} = -\frac{1}{2}\,\Box\,\frak{h}.
\end{eqnarray}

Therefore, from the Ricci tensor one obtains the unique natural generalization of the linearized equation (\ref{linearized}) as:
\begin{eqnarray}\label{X}
\frak{Ric}
=
8\pi G \left(\frak{T}-\frac{1}{2}\frak{g}\,T\right) \quad \mbox{with} \quad {\bf u}=\left(\frac{dt}{ds}, \frac{d{\bf x}}{ds} \right),
\end{eqnarray}
which coincides with the form of the field equation that Einstein presented in 1915.

 This is a consequence of Lovelock’s theorem \cite{Lovelock}, which states that if we want a symmetric tensor constructed locally from second-order derivatives of the first fundamental form $\frak{g}$
and that automatically guarantees the conservation of the energy-momentum tensor, the only option is the combination $\frak{Ric}+\Lambda\frak{g}$
 and $\frak{g}T$, as it appears in Einstein’s original 1915 version (\ref{X}) with $\Lambda=0$, which he later also used to describe his cosmological model \cite{Einstein1917}.

}

\section{
Beyond Geometry: On the Ontology of Gravitation and Spacetime}

In this final section, we address in some detail the traditional interpretation of gravitation in geometric terms. Since Einstein’s formulation of General Relativity, the curvature of spacetime has been regarded as the manifestation of the gravitational field itself. Yet such an interpretation raises profound conceptual questions: if spacetime is not a material medium, in what sense can it be said to ``curve''? And if the metric field admits infinitely many mathematically distinct representations that are physically equivalent, as the hole argument reveals, what is the ontological status of this curvature?

Here, we adopt the standpoint outlined in the previous section. Gravitation is not conceived as a geometric deformation of a physical substrate, but rather as a dynamic relation between matter, inertia, and the field that mediates their interaction. Geometry, in this view, functions as a language—a mathematical structure encoding the dynamical relations among physical entities, rather than an entity possessing its own physical essence. 

This relational reading brings us closer to Mach's spirit and to Einstein’s mature reflections, particularly those of his 1920 Leiden address and of his later writings, where the notion of an ``ether'' regains meaning—not as a substance filling space, but as the expression of the physical conditions that determine metric relations.

We shall first examine the Machian foundations underlying this interpretation, and then consider how Einstein’s evolving concept of the ether can be reinterpreted within this relational framework.

\subsection{A Machian Interpretation of Gravitation}

In the spirit of Mach’s critique of absolute space, gravitation may be understood as a manifestation of the dynamical relation between matter and inertia, rather than as an autonomous curvature of spacetime. In this view, the fundamental invariant \( ds \)—the infinitesimal proper time in the Minkowskian framework—does not represent an independent geometrical entity, but a relational descriptor encoding how clocks, rods, and trajectories are influenced by the surrounding distribution of matter and energy. It expresses the local structure of inertia shaped by the presence of mass and energy, rather than a background geometry, which exists by itself.

From this standpoint, matter defines the inertial properties of space and simultaneously generates the potential that modifies the Minkowski invariant. Gravitation thus appears as a dynamical alteration of the invariant of Special Relativity, not as an intrinsic curvature of an empty manifold. When the sources vanish, the invariant naturally reverts to its Minkowski form, corresponding to pure inertial motion in the absence of gravitational influence.

In contrast with General Relativity, where the field equations admit non-trivial vacuum solutions ($\frak{T}=0$) describing self-sustained geometrical configurations, the present approach admits only the Minkowski invariant in the absence of matter. In Einstein’s formulation, geometry acquires a degree of autonomy: spacetime might remain curved even without material sources. By contrast, in the present framework the inertial structure—represented by the invariant $ds^2$—exists only as a relational property determined by the distribution of matter and energy. Once matter is removed, the invariant necessarily returns to its Minkowski form, restoring pure inertia.

Within this framework, inertia and gravitation emerge as two aspects of a single dynamical structure. The invariant that defines the proper time of free motion in the absence of matter is not an immutable background quantity but the very object modified by the presence of mass-energy. Through this modification, matter determines both the inertial structure and the gravitational interaction. In this sense, the Equivalence Principle acquires a genuinely dynamical meaning: the same invariant that governs free motion in the absence of gravitation becomes the generator of gravitational effects when matter is present. The absence of matter leaves only the Minkowski invariant, ensuring that inertia is entirely determined by the material content of the universe.

\subsection{The Ether and the Relational Nature of Spacetime}

In his 1920 Leiden lecture \cite{Einstein1920}, Einstein argued that the general theory of relativity restores the concept of ether, albeit in a completely new form. This ether is not a mechanical medium nor a material substrate for forces, but a continuous entity that determines the metric properties of spacetime and conditions the propagation of physical phenomena. As Einstein emphasized, there can be no space devoid of the gravitational field, since it is precisely this field that confers to space its physical structure. Spacetime thus appears as a medium endowed with a state, though devoid of any material substance.

Nevertheless, this formulation still retains a certain substantive flavor, insofar as it seems to attribute to the metric field a physical reality of its own.

In contrast to Einstein’s Leiden interpretation and to Lehmkuhl’s \cite{Lehmkuhl} analysis, the present approach goes one step further. The metric field $\frak{g}$ is not conceived as the physical state of a medium—even a non-material one—but as a symbolic representation of the dynamical relations between matter, inertia, and proper time. The Hole Argument makes this point explicit: infinitely many mathematically distinct metrics can correspond to the same physical situation, which excludes the possibility of attributing to the metric tensor any ontological status. What remains invariant across these representations is the relational structure that governs motion—the balance between gravitational and inertial effects expressed through the proper time. In this sense, spacetime geometry is not a physical fabric nor a field with independent reality, but a mathematical codification of the dynamical organization of matter and inertia within the universe.

From this viewpoint, geometry is not an essential reality, but just a descriptive language expressing the relational dynamics of the gravitational field. Spacetime is not a physical fabric that bends, but a representational structure reflecting how gravitation and inertia are organized into a coherent system of relations. These relations are manifested through the proper time, which governs the motion of test particles and encodes the local balance between gravitational and inertial effects. The invariant element \(ds\) thus acquires a concrete dynamical meaning: it represents the infinitesimal lapse of proper time that determines the law of free motion, rather than an abstract geometrical quantity of a pseudo-Riemannian manifold. 

Einstein himself tended to regard this invariant primarily as a geometrical measure of spacetime intervals, rather than as a dynamical principle governing motion. His reluctance to interpret \(ds\) as a physical proper-time reflects his commitment to a geometric ontology of gravitation, in which the metric expresses the structure of spacetime rather than the evolution of matter–inertia relations. From a relational perspective, by contrast, the extremization of proper time provides the true dynamical law: when the balance between gravitation and inertia is exact, as in free fall, no physical medium or external force needs to be invoked—the motion follows solely from the principle of proper time.

In the last years of his life, Einstein moved decisively in this direction. To wit, in his later reflections, he acknowledged that physical reality does not consist in geometry, but rather in the field that geometry describes. In a 1952 letter to Lincoln Barnett, he wrote \cite{Einstein1952}:

\begin{quote}
“{\it The field is the only reality. The material things are nothing but regions of the field where the field is extremely intense. There is no place in this new kind of physics both for the field and the matter, for the field is the only reality. Physical reality is not geometry itself, but the field which the geometry describes.}”
\end{quote}

However, this conviction was not entirely new. Already in 1948, in a letter to Barnett, Einstein had explicitly rejected the idea that general relativity “geometrizes” gravitation:

\begin{quote}
“{\it I do not agree with the idea that the general theory of relativity is geometrizing physics or the gravitational field. The concepts of physics have always been geometrical concepts, and I cannot see why the \( \frak{g} \) field should be called more geometrical than, for instance, the electromagnetic field or the distances of bodies in Newtonian mechanics. The notion probably arises from the fact that the mathematical origin of the \( \frak{g} \) field lies in the Gauss–Riemann theory of the metric continuum, which we are accustomed to regard as part of geometry. I am convinced, however, that the distinction between geometrical and other kinds of fields is not logically founded.}”
\end{quote}

These words confirm that, for Einstein, the metric field was never a geometrical substance but a physical representation of dynamical relations. In this sense, his mature view converges with the relational interpretation developed here: geometry is not an ontological reality, but a mathematical expression of how matter and inertia interact through proper time.

{

}



{

\section{Conclusions}

The development of General Relativity incarnates a profound interplay between physical principles and mathematical formalism. From Minkowski’s unification of space and time, through Planck’s early variational insight, to Einstein’s 1907–1912 attempts to incorporate gravitation via the invariant $ds^2$, the guiding thread was always physics: the Equivalence Principle, energy–momentum conservation, and the Principle of Relativity. Geometry entered decisively only in 1913, when Grossmann introduced Einstein to Riemannian methods, culminating in the 1915 field equations. One must accept that without these advanced mathematical tools the equations of General Relativity could not have been formulated, but the price paid at the moment was probably too high: moving the emphasis from physics to mathematics. And this view has remained up to now.

The prevailing interpretation treats gravitation geometrically: spacetime as a pseudo-Riemannian manifold, curvature replacing the Newtonian force, free particles following geodesics, and matter–energy shaping geometry. Yet, as the hole argument shows, treating the metric as a physical substance threatens determinism. Einstein’s resolution—that diffeomorphic solutions represent the same physical situation—already hints at a relational understanding of spacetime. Later developments, including teleparallel gravity and Einstein’s 1920 notion of a non-mechanical “ether,” reinforce the idea that geometry serves primarily as a descriptive framework rather than an ontological entity.

This work emphasizes a complementary perspective that Einstein himself explored before 1913: a formulation in which the invariant $ds^2$ is dynamically shaped by matter through the Equivalence Principle and Lorentz invariance, without invoking curvature as a fundamental postulate. In this understanding, $ds$ and its integral along a worldline, $s = \int ds$, play the central role. In Minkowski’s Special Relativity, $s$ represents the proper time along a trajectory; in the presence of gravitation, the proper time is modified so that free-falling particles follow extremal paths of $s$. Gravitation thus appears as a modification of the flow of proper time, governing both motion and the measurement of time.

From this perspective, curvature or torsion are auxiliary tools: the fundamental invariant encodes how proper time is locally measured, while the consistency of the dynamical description—how particles move, clocks measure time, and fields propagate—remains what matters physically. Einstein himself warned against reifying spacetime; the metric and its invariants are not physical entities, but relational constructs governing motion.

In the weak-field regime, this framework naturally recovers the Newtonian limit, including velocity-dependent corrections, while avoiding the conceptual difficulties highlighted by the hole argument. Here, the invariant $ds$—interpreted as the differential proper-time—acts as a dynamical expression of the interaction between matter and inertia, rather than as the metric of an autonomous geometrical entity. In this sense, this view continues and clarifies Einstein’s early insight: that gravitation can be described as a modification of inertial structure, without involving spacetime curvature.

It is important to recognize that the relational and dynamical interpretation of gravitation has historically been met with skepticism. Many physicists, especially after the 1930s, considered these ideas largely philosophical, with limited perceived empirical relevance. Subsequent generations concentrated on quantum theory and particle physics, leaving General Relativity—and particularly its conceptual foundations—on the periphery. This was helped by the very important fact that, with the arrival of quantum physics, the concept itself of physical reality has been questioned. The emphasis has eventually moved to the mathematical models that try to approach the inaccessible ``physical reality", to describe it in the best possible way \cite{EE_book21}.

Anyway, the perspective above for the case of General Relativity helps to clarify the deep relation between matter, inertia, and gravitation, while fully preserving all practical predictions of the conventional geometric formulation.

In conclusion, General Relativity in its weak-field limit admits two complementary readings: as a geometric theory of curved spacetime, and as a dynamical framework describing the interaction between matter and inertia through the invariant $ds$. While the geometric interpretation remains elegant and powerful, it does not exhaust the physical content of the theory. By recovering linearized General Relativity in the harmonic gauge, preserving its empirical successes, and clarifying the operational meaning of the Equivalence Principle, the perspective developed here restores the physical foundation of the theory and provides a conceptually transparent understanding of gravitation in its weak-field domain. {Moreover, when combined with the structure of the Ricci tensor and the uniqueness results embodied in Lovelock’s theorem, this analysis indicates that any consistent extension to stronger gravitational fields must necessarily be formulated in terms of the Ricci tensor.}

Ultimately, the essence of gravitation lies not in mathematical geometry, but in the dynamical coherence of matter, inertia, and proper time. Geometry serves as the symbolic expression of this coherence—a language through which the universe reveals its relational order. From this perspective, Einstein’s quest for unity between inertia and gravitation regains its original meaning through the principle of extremal proper time, which embodies the mutual shaping of matter and motion within a single dynamical law.

}

\begin{acknowledgments}

JdH is supported by the Spanish grant PID2021-123903NB-I00
funded by MCIN/AEI/10.13039/501100011033 and by ``ERDF A way of making Europe''. EE is partly supported by the program Unidad de Excelencia María de Maeztu, CEX2020-001058-M, and by AGAUR, project 2021-SGR-00171.

\end{acknowledgments}

\end{document}